\documentclass[10pt,conference]{IEEEtran}
\IEEEoverridecommandlockouts

\usepackage{graphicx}
\usepackage{textcomp}
\usepackage{enumitem}
\usepackage{multirow}
\usepackage{colortbl}
\usepackage{array}
\usepackage{tcolorbox}
\usepackage{nicematrix}
\usepackage{tabularx}
\usepackage{balance}
\usepackage{colortbl}
\usepackage{booktabs}
\usepackage{url}
\usepackage{calc}
\usepackage{siunitx}
\usepackage{makecell}

\newcommand{\pval}[2]{%
  \ifnum#1<50 \textbf{\small (#2)} \else \small (#2) \fi
}

\definecolor{lightgray}{gray}{0.9}

\PassOptionsToPackage{normalem}{ulem}
\usepackage{ulem}

\providecolor{added}{rgb}{0,0,1}
\providecolor{deleted}{rgb}{1,0,0}

\usepackage{subcaption}
\usepackage{mwe}
\usepackage{hyperref}

\hypersetup{
    colorlinks=false,
    pdfborder={0 0 0}
}

\definecolor{specialblue}{RGB}{0,104,149}
\definecolor{specialgray}{RGB}{242,242,242}

\usepackage{gb4e}
\noautomath

\usepackage{tikz}
\usepackage[framemethod=tikz]{mdframed}
\usetikzlibrary{calc}

\newtcolorbox{mybox}{
  sharp corners,
  colback=specialgray,
  colframe=specialblue,
  boxrule=0pt,
    toprule=0pt,
  bottomrule=0pt,
  leftrule=3pt, 
  rightrule=3pt 
}

\usepackage{cite}
\usepackage{amsmath,amssymb,amsfonts}
\usepackage{algorithmic}
\def\BibTeX{{\rm B\kern-.05em{\sc i\kern-.025em b}\kern-.08em
    T\kern-.1667em\lower.7ex\hbox{E}\kern-.125emX}}

\usepackage{eso-pic}

\makeatletter
\AddToShipoutPictureBG*{%
    \AtPageLowerLeft{%
        \ifnum\thepage=1
            \put(\dimexpr(\paperwidth-\textwidth)/2\relax+8pt,35){
                \color{black}
                \makebox[\textwidth-25pt][c]{
                    \setlength{\fboxsep}{3pt}  
                    \fbox{\parbox[c]{\dimexpr\textwidth-0\fboxsep-0\fboxrule\relax}{\centering \footnotesize \textsf{© 2025 IEEE. Personal use of this material is permitted. Permission from IEEE must be obtained for all other uses, in any current or future media, including reprinting/republishing this material for advertising or promotional purposes, creating new collective works, for resale or redistribution to servers or lists, or reuse of any copyrighted component of this work in other works.}}}
                }
            }
        \fi
    }
}
\makeatother

\begin{document}

\title{GUIDE: LLM-Driven GUI Generation Decomposition for Automated Prototyping}



\author{
\IEEEauthorblockN{
Kristian Kolthoff\IEEEauthorrefmark{2}\IEEEauthorrefmark{1}, 
Felix Kretzer\IEEEauthorrefmark{3}\IEEEauthorrefmark{1}, 
Christian Bartelt\IEEEauthorrefmark{2}, 
Alexander Maedche\IEEEauthorrefmark{3},
and Simone Paolo Ponzetto\IEEEauthorrefmark{4}}

\IEEEauthorblockA{\IEEEauthorrefmark{1}Authors contributed equally to the paper.}

\IEEEauthorblockA{
\IEEEauthorrefmark{2}Institute for Software and Systems Engineering,
Clausthal University of Technology, Clausthal, Germany\\
Email: \{kristian.kolthoff, christian.bartelt\} @tu-clausthal.de}

\IEEEauthorblockA{\IEEEauthorrefmark{3}Human-Centered Systems Lab,
Karlsruhe Institute of Technology, Karlsruhe, Germany\\
Email: \{felix.kretzer, alexander.maedche\} @kit.edu}

\IEEEauthorblockA{\IEEEauthorrefmark{4}Data and Web Science Group, 
University of Mannheim, Mannheim, Germany\\
Email: simone@informatik.uni-mannheim.de}}



\maketitle

\begin{abstract}
Graphical user interface (GUI) prototyping serves as one of the most valuable techniques for enhancing the elicitation of requirements, facilitating the visualization and refinement of customer needs and closely integrating the customer into the development activities. While GUI prototyping has a positive impact on the software development process, it simultaneously demands significant effort and resources. The emergence of Large Language Models (LLMs) with their impressive code generation capabilities offers a promising approach for automating GUI prototyping. Despite their potential, there is a gap between current LLM-based prototyping solutions and traditional user-based GUI prototyping approaches which provide visual representations of the GUI prototypes and direct editing functionality. In contrast, LLMs and related generative approaches merely produce text sequences or non-editable image output, which lacks both mentioned aspects and therefore impede supporting GUI prototyping. Moreover, minor changes requested by the user typically lead to an inefficient regeneration of the entire GUI prototype when using LLMs directly. In this work, we propose \textit{GUIDE}, a novel LLM-driven GUI generation decomposition approach seamlessly integrated into the popular prototyping framework \textit{Figma}. Our approach initially decomposes high-level GUI descriptions into fine-granular GUI requirements, which are subsequently translated into \textit{Material Design} GUI prototypes, enabling higher controllability and more efficient adaption of changes. To efficiently conduct prompting-based generation of \textit{Material Design} GUI prototypes, we propose a retrieval-augmented generation (RAG) approach to integrate the component library. Our preliminary evaluation demonstrates the effectiveness of \textit{GUIDE} in bridging the gap between LLM generation capabilities and traditional GUI prototyping workflows, offering a more effective and controlled user-based approach to LLM-driven GUI prototyping. Video presentation of \textit{GUIDE} is available at: \url{https://youtu.be/C9RbhMxqpTU}
\end{abstract}


\IEEEpeerreviewmaketitle

\section{Introduction}

Over the years, a diverse collection of methods have been introduced and implemented to enhance the process of requirements elicitation in the development of user-centric software systems \cite{pohl2010requirements}. Among these methods, GUI prototyping has emerged as a particularly potent approach, since it equips analysts with the means to visually articulate their understanding of the requirements, while simultaneously providing customers with a tangible artifact for validation. Moreover, prototypes act as a catalyst for engaging customers throughout the development process, fostering insightful dialogues that can lead to the clarification and refinement of requirements \cite{ravid2000method, beaudouin2002prototyping}. However, creating high-fidelity GUI prototypes is a resource intensive activity, both in time and cost.

To enhance traditional GUI prototyping and increase the level of automation, recent research focused on exploiting the generative capabilities of LLMs. For example, \textit{Instigator} \cite{brie2023evaluating} is a \textit{GPT} model trained on extensive web data, to generate low-fidelity GUI layouts from text descriptions and chosen component types. An alternative approach is to fine-tune an existing pretrained language model \cite{feng2023designing}. Moreover, \textit{MAxPrototyper} \cite{yuan2024maxprototyper} supports the generation of high-fidelity GUI prototypes through prompting LLMs. However, this approach necessitates providing both text descriptions and a fully designed GUI layout as input and subsequently outputs GUI prototypes in an unique domain-specific language (DSL). While the mentioned research exploits LLMs to directly generate GUI prototypes, these prototypes are represented textually, typically using a DSL. In addition, LLMs have also shown to possess tremendous code generation capabilities \cite{chen2021evaluating, kolthoff2024zero, fiebig2025effective}, including languages used for GUI prototypes such as HTML/CSS. 

However, there is currently a gap between LLMs generating GUI code as text \cite{kolthoff2024zero, fiebig2025effective} and established  GUI prototyping workflows, namely visual editors such as \textit{Figma} \cite{figma_tool} widely employed by practitioners \cite{kretzer2023}. Specifically, because LLMs are primarily designed for text generation, their outputs are not easily visualizable, posing a challenge for their deep integration in GUI prototyping. Furthermore, modifications to GUI prototypes are generally performed by the prototype developers visually rather than through direct code manipulation. Moreover, when minor changes of generated GUI prototypes are requested by the user, LLMs typically inefficiently regenerate the entire GUI prototype. 

\begin{figure*}[!t]
  \centering
 \includegraphics[width=0.82\textwidth]{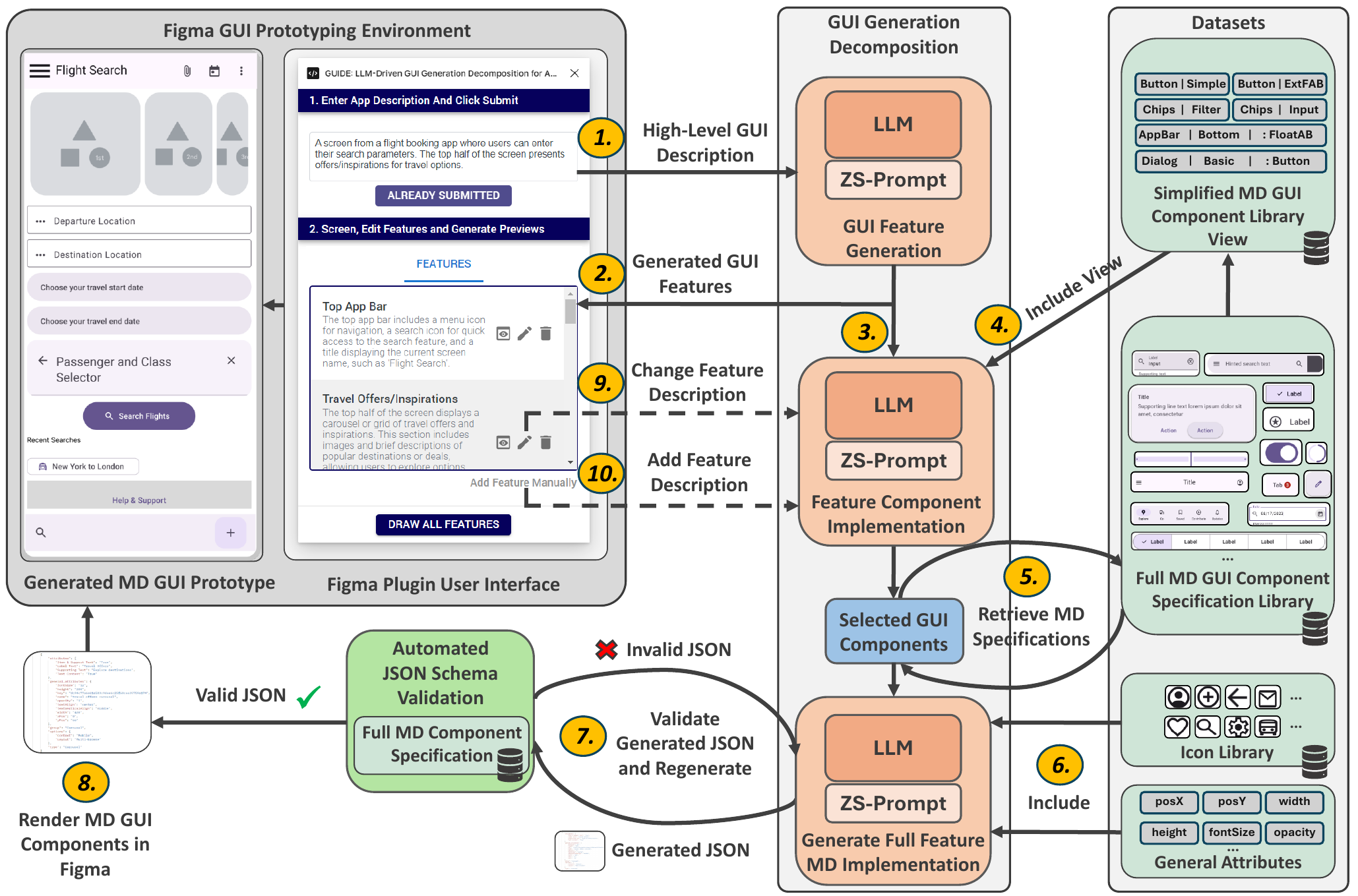}
  \caption{Overview of the \textit{GUIDE} architecture with \textit{Figma} plugin, GUI generation decomposition and \textit{Material Design} datasets}
	\label{fig:overview}
 \vspace{-6mm}
\end{figure*}

To close this gap, in this work, we introduce \textit{GUIDE}, a novel LLM-driven GUI generation decomposition approach which is seamlessly integrated into the popular prototyping tool \textit{Figma}. Our approach derives low-level GUI features from high-level text descriptions provided by users and directly generates a corresponding editable GUI prototype using the popular \textit{Material Design} component library \cite{material_design}. Moreover, \textit{GUIDE} facilitates the efficient modification of existing features or the addition of new ones within the GUI prototype, eliminating the need to regenerate the entire prototype. Thus, this integration combines the LLM generation capabilities with the advantages of traditional visual GUI prototyping approaches, enabling both more effective and efficient GUI prototyping. Our approach can help prototype developers as well as analysts to rapidly and effectively create prototypes based on high-level textual requirements. Our source code, datasets and prototype are available at our repository \cite{github}.
\section{Approach: GUIDE}

\textit{GUIDE} is composed of two main components including \textit{(A)} the \textit{Figma} plugin enabling users to interact with our approach via an easy-to-use interface and \textit{(B)} the GUI generation decomposition, which is an LLM-based approach for \textit{(i)} decomposing the high-level GUI description from the user into a fine-granular GUI feature collection, \textit{(ii)} selecting relevant combinations of GUI component types for each GUI feature from a prespecified GUI component library and \textit{(iii)} generating the fully specified GUI component implementations. An overview of the \textit{GUIDE} architecture is provided in Figure \ref{fig:overview}.

\subsection{Figma Plugin User Interface}

Initially, prototype developers can provide their concise and high-level GUI descriptions for the prototypes they are assigned to create directly into the user interface of \textit{GUIDE}, which is implemented as a plugin within \textit{Figma}. Subsequently, \textit{GUIDE} will derive a fine-granular feature collection displayed in the plugin with a name and a short textual description of the feature. Our approach facilitates users to rapidly screen the generated feature list, directly edit the feature description, delete or add additional features manually. Moreover, \textit{GUIDE} enables users to directly render the implementation of the suggested features as editable GUI components within \textit{Figma}. 

\subsection{GUI Generation Decomposition}

To improve the generated GUI prototypes and enhance control over the generation process, we decompose the GUI generation task into multiple smaller steps \cite{khot2022decomposed}. As research has shown before, this can not only improve the effectiveness of the LLM in completing complex tasks, but also follows a more human-like approach to creating GUI prototypes \cite{khot2022decomposed}. 

\subsubsection{GUI Feature Generation}

Based on the high-level GUI description provided by the user, we employ an LLM in a zero-shot prompting approach \cite{radford2019language} to generate the corresponding GUI feature collection. In particular, we instruct the LLM to focus solely on recommending functional features that can be represented visually in a GUI prototype, therefore neglecting non-functional requirements (e.g., responsive design, accessibility). In addition, we provide a JSON schema to the LLM including a name and short description of the requirements and instruct the model to directly generate the respective JSON format. This representation can then easily be consumed by the front-end and by the subsequent computation steps.

\subsubsection{Retrieval-Augmented GUI Feature Implementation}

After generating a feature collection, we decompose the subsequent generation of the respective feature implementations as GUI components within \textit{Figma} into two separate steps. Within \textit{GUIDE}, we integrated the popular \textit{Material Design (MD)} component library \cite{material_design} to enable the generation of respective GUI prototypes. In particular, the library consists of over 59 different components, ranging from simple individual components (e.g., \textit{Button}, \textit{Checkbox}, \textit{Label}) to more complex component groups (e.g., \textit{Search Bar}, \textit{Dialog}, \textit{Top App Bar}). Although LLMs are trained on massive amounts of textual data, the specifically employed component library with the particular configuration options and syntax cannot directly be generated by the LLM. Since the full JSON specification of each GUI component encompasses many different aspects and typically only a small amount of components are relevant for implementing a feature, incorporating the entire specification into the GUI generation as part of a zero-shot prompt for an LLM would be inefficient. Therefore, we propose a retrieval-augmented generation (RAG) \cite{lewis2020retrieval} approach for the implementation of GUI features. In particular, we first automatically derive a simplified component library view from the full specification library (only including the component group (e.g., \textit{Button}) and component type e.g., \textit{FloatingActionButton} information), significantly reducing the required tokens (reduction of approximately 60\% of consumed tokens). Afterwards, we construct a zero-shot prompt that incorporates the simplified library in the context and instructs the LLM to select the relevant component types to implement the respective GUI feature. Subsequently, we employ the selected GUI components to retrieve the full JSON specifications of the relevant GUI components and utilize them in a second zero-shot prompt that finally generates the actual GUI feature implementation with \textit{MD} components. In addition, we provide an icon collection and a set of general attributes (e.g., \textit{posX}, \textit{posY}, \textit{width}, \textit{height}), that each component possesses, into the generation process. The LLM is instructed to generate the full implementation for each GUI feature using a prespecified JSON format, which can be used to render the components in \textit{Figma}. To improve the reliability of the JSON generation, we additionally conduct an automated validation of the generated JSON format by comparing it to the respective JSON schema.

\subsection{Prototype Implementation}

The \textit{GUIDE} prototype is implemented as a \textit{Figma} plugin using \textit{TypeScript} for the front-end. For enabling the integration of the LLM into generating prototypes in \textit{Figma}, we created a proprietary JSON format that can be interpreted both by the LLM and the plugin for rendering. The back-end of \textit{GUIDE} is implemented as a Python-based \textit{Quart} app running within a highly scalable \textit{gunicorn}/\textit{nginx} server providing a REST API for the described functionality. For the LLM, we employed the most recent \textit{GPT-4o} model from \textit{OpenAI} (128k tokens context length with a maximum of 4,095  output tokens, accessed in October 2024), representing the multi-model extension of their prior \textit{GPT-4} model with state-of-the-art performance \cite{openai2023gpt4}.
\section{Experimental Evaluation}
In the following, we describe the setup of our preliminary evaluation. Our evaluation consists of a small between-subjects lab-based study in which participants created GUI prototypes with (treatment) and without (control) our assistant based on app descriptions. These GUI prototypes were later evaluated by several crowd-workers with UI/UX experience on \textit{Prolific}.

\subsection{GUI Prototype Generation}
\textit{Procedure.}
Participants started the first sub-study with a questionnaire and watching an initial video introduction on the prototyping tool, including the component library \textit{Material Design 3} \cite{material_design}. In addition, the treatment group watched an introduction to the use of our assistant, including the core functions: \textit{(i)} letting the assistant create GUI features based on the short description, \textit{(ii)} editing existing features and adding custom features, \textit{(iii)} previewing the generation, and \textit{(iv)} drawing the generation in the prototyping tool. The participants then conducted a 45-minutes GUI prototyping phase and were instructed to create four GUI prototypes for four short descriptions. After the prototyping phase, the treatment group answered questions about using the assistant in the survey.

\vspace{0.10cm}
\textit{Participants.}
We recruited 11 participants (3 female, 8 male) from a university pool to create GUI prototypes in a lab-based environment. On average, participants were 26.64 years old ($\sigma=5.68$) and studied for 5.05 years ($\sigma=1.44$). Participants were randomly assigned to either the control group (six participants) or treatment group (five participants). 

\begin{table*}[]
\caption{Median, mean, and results of Wilcoxon rank sum tests (two-sided) for GUIs ratings on Likert scales, (Strongly Disagree (1) to Strongly Agree (9)) generated for GUIs from the control and treatment groups. Crowd-workers on \textit{Prolific} rated GUIs: \textit{meets requirements from description}, \textit{utilizes correct components}, \textit{texts fit to GUI description}, \textit{design consistent with GUI description}, \textit{design appealing}, \textit{clear information organization}, \textit{intuitive interactions}, \textit{minimal errors}, and \textit{overall satisfaction}.}
\label{tab:results_prolific}
\renewcommand{\arraystretch}{1.10}
\small
\centering
\begin{tabular}{|l||l||c|c|c|c|c|c|c|c|c|}
\hline
                        \rowcolor{gray!30}   & \textbf{Measure} & \textbf{Requi.}    & \textbf{Comp.}    & \textbf{Text}    & \textbf{Consist.}    & \textbf{Appeal.}    & \textbf{Inf. Org.}    & \textbf{Intui.}    & \textbf{Errors}    & \textbf{Satisf.}    \\ \hline  \hline

\multirow{2}{*}{\textbf{Contr.}}   & Median &\scriptsize 5     &\scriptsize 5     &\scriptsize 6     & \scriptsize 5     & \scriptsize 4     & \scriptsize 6     & \scriptsize 5     & \scriptsize 4     & \scriptsize 3     \\ \cline{2-11} 
                          & \cellcolor{lightgray!60}Mean   & \cellcolor{lightgray!60}\scriptsize 4.746 & \cellcolor{lightgray!60}\scriptsize 4.769 & \cellcolor{lightgray!60}\scriptsize 5.015 & \cellcolor{lightgray!60}\scriptsize 5.000 & \cellcolor{lightgray!60}\scriptsize 4.354 & \cellcolor{lightgray!60}\scriptsize 5.154 & \cellcolor{lightgray!60}\scriptsize 4.762 & \cellcolor{lightgray!60}\scriptsize 4.392 & \cellcolor{lightgray!60}\scriptsize 4.015 \\ \hline  \hline
\multirow{2}{*}{\textbf{Treat.}} & Median &  \scriptsize 8     & \scriptsize 7     & \scriptsize 7     & 
 \scriptsize7     &  \scriptsize 7     & \scriptsize 7     &  \scriptsize7     & \scriptsize 7     & \scriptsize 6.5   \\ \cline{2-11} 
                           & \cellcolor{lightgray!60}Mean   & \cellcolor{lightgray!60}\scriptsize 7.393 &  \cellcolor{lightgray!60}\scriptsize 7.107 & \cellcolor{lightgray!60}\scriptsize 6.713 & \cellcolor{lightgray!60}\scriptsize 7.260 & \cellcolor{lightgray!60}\scriptsize 6.127 & \cellcolor{lightgray!60}\scriptsize 6.973 & \cellcolor{lightgray!60}\scriptsize 6.967 & \cellcolor{lightgray!60}\scriptsize6.227 & \cellcolor{lightgray!60}\scriptsize 6.187 \\ \hline  \hline
\multirow{3}{*}{\begin{tabular}[c]{@{}l@{}}\textbf{Wilcox.} \\ \textbf{Rank} \\ \textbf{Sum} \end{tabular}} &
  p-Value &
\scriptsize$7.5e{-20}$ &
\scriptsize$3.6e{-15}$ &
\scriptsize$5.6e{-10}$ &
\scriptsize$9.9e{-16}$ &
\scriptsize$3.1e{-10}$ &
\scriptsize$3.0e{-11}$ &
\scriptsize$8.6e{-15}$ &
\scriptsize$7.1e{-10}$ &
\scriptsize$5.2e{-13}$ \\ \cline{2-11}

 &
 \cellcolor{lightgray!60}Conf. Int. &
  \cellcolor{lightgray!60}\scriptsize $[-3, -2]$ &
  \cellcolor{lightgray!60}\scriptsize $[-3, -2]$ &
  \cellcolor{lightgray!60}\scriptsize $[-2, -1]$ &
  \cellcolor{lightgray!60}\scriptsize $[-3, -2]$ &
  \cellcolor{lightgray!60}\scriptsize $[-2, -1]$ &
  \cellcolor{lightgray!60}\scriptsize $[-2, -1]$ &
 \cellcolor{lightgray!60} \scriptsize $[-3, -2]$ &
 \cellcolor{lightgray!60} \scriptsize $[-3, -1]$ &
  \cellcolor{lightgray!60}\scriptsize $[-3, -2]$ \\ \cline{2-11} 
   & r-Value &
  \scriptsize 0.545 &
  \scriptsize 0.470 &
  \scriptsize 0.371 &
  \scriptsize 0.480 &
  \scriptsize 0.376 &
  \scriptsize 0.397 &
  \scriptsize 0.464 &
  \scriptsize 0.368 &
  \scriptsize 0.431 \\ \hline
\end{tabular}
\vspace{-0.2cm}
\end{table*}

\subsection{GUI Prototype Evaluation}
\textit{Procedure.}
To evaluate the GUI prototypes created in our experiment, we used crowd-workers with UI/UX experience and conducted an online experiment. In a survey, the crowd-workers evaluated ten randomly drawn prototypes after agreeing to the experiment and data processing.

\vspace{0.10cm}
\textit{Participants.}
We invited 30 participants through \textit{Prolific} with UI/UX design experience, a high acceptance rate on \textit{Prolific} and certain language skills (English and German, since some GUI prototypes contained German texts). We excluded two participants for failing at least one attention check. Hence, we eventually considered data from 28 participants (12 female, 1 diverse, 15 male). Participants had an average age of 33.21 years ($\sigma=8.23$), 7.36 ($\sigma=7.71$) years of experience creating design and 6.46 ($\sigma=6.97$) years experience in judging design. On average, participants spent 16 minutes on our survey.

\subsection{Evaluation Results}
Subsequently, we briefly present the results of both sub-studies. 
Although both groups in the lab-based sub-study had 45 minutes to create the GUI prototypes and identical short GUI descriptions, participants started to work on significantly more GUI prototypes using \textit{GUIDE} (3.2 per participant, 16 total with five participants) than in the control group (2.33 per participant, 14 total with six participants).
Table \ref{tab:results_prolific} shows the medians, mean values, and the results of Wilcoxon rank sum tests for the obtained crowd-worker ratings. For each of the asked questions, the mean and median ratings were higher for the treatment group. The Wilcoxon rank sum tests (two-sided) were significant for each of the evaluated questions, indicating the high effectiveness of our plug-in and our overall approach.
\section{Related Work}

In the realm of GUI prototyping, recent methodologies have increasingly leveraged LLMs to enhance the generation of GUI prototypes with varying levels of detail. For instance, the \textit{Instigator} approach \cite{brie2023evaluating} involves training a complete minified \textit{GPT} model from the ground up. This method utilizes a vast repository of automatically collected web pages, which are subsequently converted into low-fidelity layouts for training purposes. In contrast, another strategy involves fine-tuning a pre-existing LLM to generate GUIs by utilizing the \textit{Rico} GUI repository \cite{feng2023designing}. Although both methods are capable of creating GUI layouts or prototypes from concise textual descriptions, they are hindered by the need for substantial training resources, the production of low-fidelity prototypes, and the reliance on domain-specific languages (DSL) for representing GUI prototypes, which restricts their overall applicability. Similarly, the \textit{MAxPrototyper} \cite{yuan2024maxprototyper} employs an LLM to generate a custom DSL for GUI prototypes based on brief text descriptions and predefined GUI layouts. However, this approach primarily concentrates on generating GUI content, such as text and images, that aligns with the given text description. Meanwhile, recent studies on zero-shot prompting for GUI prototyping have been limited to verifying the implementation of requirements within GUI prototypes \cite{kolthoff2024interlinking}, without exploring the potential of zero-shot prompting for the comprehensive generation of GUI prototypes from descriptions to support users. In contrast to the mentioned methods, our approach \textit{GUIDE} sets itself apart by seamlessly combining the generative abilities of LLMs with traditional GUI prototyping tools such as \textit{Figma}, empowering prototype developers and enabling rapid creation of effective GUI prototypes. In addition, our RAG approach ensures the efficient generation of respective GUI feature implementations.

\section{Conclusion \& Future Work}

In this paper, we proposed \textit{GUIDE}, an approach leveraging the zero-shot GUI generation capabilities of recent LLMs with the advantages of traditional GUI prototyping tools such as \textit{Figma}. This integration facilitates the rapid creation of effective GUI prototypes and provides an effective support for prototype developers. For future work, we plan to integrate the recommendation of multiple implementation variants for a GUI feature, providing additional assistance to practitioners.

\ifCLASSOPTIONcaptionsoff
  \newpage
\fi

\bibliographystyle{IEEEtran}
\bibliography{bibtex/bib/ref}

\end{document}